\documentclass[english,aps,prl,showpacs,amsmath,amssymb,twocolumn]{revtex4}
\usepackage[T1]{fontenc}
\usepackage[latin1]{inputenc}
\usepackage{graphicx}
\makeatletter

\usepackage{epsfig}

\usepackage{dcolumn}
\usepackage{bm}
\usepackage{epsf}

\usepackage{babel}
\makeatother
\begin{document}

\title{A single particle uncertainty relation}

\author{Thomas Sch\"urmann}

\affiliation{Planeten Strasse 25, 40223 D\"usseldorf, Germany}
\date{\today}

\begin{abstract}
We consider the successive measurement of position and momentum of a single particle. Let $\cal P$ be the conditional probability to measure the momentum $k$ with precision $\Delta k$, given a previously successful position measurement $q$ with precision $\Delta q$. Several upper bounds for the probability $\cal P$ are derived. For arbitrary, but given precisions $\Delta q$ and $\Delta k$, these bounds refer to the variation of $q$, $k$, and the state vector $\psi$ of the particle. The first bound is given by the inequality ${\cal P}\leq\frac{\Delta k\Delta q}{h}$, where $h$ is Planck's quantum of action. It is nontrivial for all measurements with $\Delta k\Delta q<h$. A sharper bound is obtained by applying the {\it Hilbert-Schmidt} norm. As our main result, the {\it least upper bound} of $\cal P$ is determined. All bounds are independent of the order with which the measuring of the position and momentum is made.
\end{abstract}


\maketitle

The measurement process in quantum mechanics plays a dual role. On one hand, it describes the way in which the state of a quantum system changes if a measurement is performed on it, thereby influencing the predictions on the future behavior of the system. On the other hand, it gives a unique prescription for the preparation of a quantum system in a definite state.
The most generally known case of this phenomenon is the complementarity between position and momentum, as expressed quantitatively in the Heisenberg uncertainty principle. Let us begin with the ordinary case of a single particle passing through a slit in a diaphragm of some experimental arrangement. Even if the momentum of the particle is completely known before it impinges on the diaphragm, the diffraction by the slit of the plane wave will imply an uncertainty in the momentum of the particle, after it has passed the diaphragm, which is the greater the narrower the slit. Now the width of the slit may be taken as the uncertainty $\Delta x$ of the position of the particle relative to the diaphragm, in a direction perpendicular to the slit. It is simply seen from de Broglie's relation between momentum and wave-length that the uncertainty $\Delta p$ of the momentum of the particle in this direction is correlated to $\Delta x$ by means of Heisenberg's general principle $\Delta x\Delta p\sim h$. In his celebrated paper \cite{H27} published in 1927, Heisenberg attempted to establish this quantitative expression as the minimum amount of unavoidable momentum disturbance caused by any position measurement. In \cite{H27} he did not give an unique definition for the 'uncertainties' $\Delta x$ and $\Delta p$, but estimated them by some plausible measure in each case separately. In \cite{H30} he emphasized his principle by the formal refinement
\begin{eqnarray}\label{H2}
\Delta x\Delta p\gtrsim h
\end{eqnarray}
However, it was Kennard \cite{K27} in 1927 who proved the well-known inequality
\begin{eqnarray}\label{K}
\sigma_x\sigma_p\geq \hbar/2
\end{eqnarray}
with $\hbar=h/2\pi$, and $\sigma_x$, $\sigma_p$ are the ordinary standard deviations of position and momentum. Heisenberg himself proved relation (\ref{K}) for Gaussian states \cite{H30}. It should be mentioned, that Kennard was the first to choose the standard deviation as a quantitative measure of uncertainty, and neither he nor Heisenberg explicitly explained why this choice should be appropriate. Thus the choice for the standard deviation was made at a very early stage in the development of quantum theory without any explicit discussion. For uncertainties represented by standard deviations, conditions ensuring their existence are less easily established, and the concept of variance is to be applied with some care. It has been pointed out that, in fact, inequality (\ref{K}) fails to express adequately the physical contents of the uncertainty principle, as summarized by expression (\ref{H2}), in case of the single-slit diffraction \cite{Be}\cite{UfHi}\cite{Uf}\cite{BGL95}. Alternative characterizations of the 'width' of a probability distribution may be defined as the length of the smallest interval which yields a given level of total probability (confidence). This concept was considered long ago in signal theory \cite{La} and took some time until it was recognized in a wider context \cite{Uf}\cite{Hi}. It is known to entail the ordinary case of variances. \\

Typically such measures analyze the degree of localizability of position and momentum distributions and refer to two separate experiments, in the sense that to each single particle either a position or a momentum measurement is applied, and the preparation is the same in both cases. Instead, Heisenberg discusses measurement processes, in which the initial preparation of the particle plays no important role. According to (\ref{H2}), position and momentum are both measured for the same particle and the key observation is that the measurement of position necessarily disturbs the particle, so that the momentum is changed by the measurement. A novel and general way expressing this degree of disturbance in a sequential measurement was recently presented by Werner \cite{W04}. Werner defines 'uncertainty' by a certain distance between probability distributions of ideal and approximate measurements. Applied to a consecutive position and momentum measurement, these uncertainties become the precision of the position measurement, and the perturbation of the conjugate variable. These precisions satisfy a measurement uncertainty relation for the trade-off between the accuracy of the position measurement and the necessary disturbance of the momentum\cite{W04}. \\

\begin{figure}[ht]
\includegraphics[width=8.0cm,keepaspectratio]{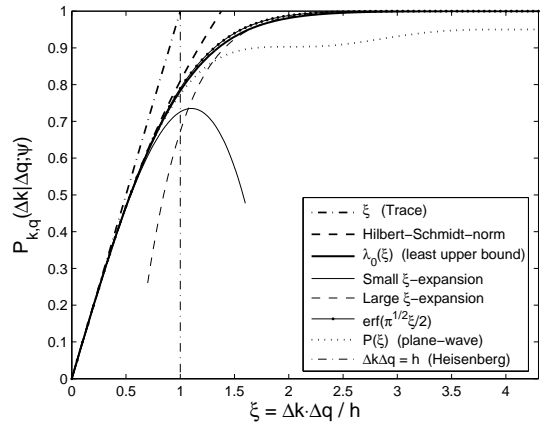}
\caption{Possible and impossible measurement probabilities (\ref{prob}). The vertical line is the dividing line of Heisenberg according to (\ref{H2}). Measuring processes with conditional probabilities above $\lambda_0(\xi)$ do not exist (see theorem).}\label{fig1}
\end{figure}

In the following we propose a similar but alternative approach. We consider the {\it conditional probability} of consecutive measurements of position and momentum. For instance, let us briefly discuss the single-slit diffraction in more detail. The slit of width $\Delta q$ provides the precision of the position measurement, and the diffraction pattern in the far-field reveal the momentum distribution. A single particle initially in a plane-wave state $\varphi(x)=1/\sqrt{\Delta x}$, of width $\Delta x>\Delta q$, will acquire a momentum spread on passing through the slit in accordance to the distribution
\begin{eqnarray}\label{psip}
|\varphi(p)|^2 =\frac{2\hbar}{\pi \Delta q} \frac{|\sin(\frac{\Delta q}{2\hbar}\,p)|^2}{p^2}
\end{eqnarray}
Then, for any precision $\Delta k$, the conditional probability to measure the particle with momentum $p\in[-\frac{\Delta k}{2},\frac{\Delta k}{2}]$ is simply computed by integrating the density (\ref{psip}). We obtain
\begin{eqnarray}\label{Prob1}
P(\xi)&=& \frac{2}{\pi}\left[\;\mbox{Si}(\pi \xi)-\frac{2}{\pi}\;\frac{\sin(\frac{\pi \xi}{2})^2}{\xi}\;\right]\\
\xi &=& \frac{\Delta k\Delta q}{h}\label{xi}
\end{eqnarray}
where $h$ is Planck's quantum of action.\footnote{The sine-integral is Si$(x)=\int_0^x\frac{\sin(t)}{t}\,dt$, \cite{AS}.}
The conditional probability (\ref{Prob1}) is explicitly dependent on the product of the precisions $\Delta k$ and $\Delta q$ (or $\xi$), ensuring the trade-off between the complementary observables. The function $P(\xi)$ is monotonically increasing, with $P(0)=0$ and $P(\xi)\to 1$ for $\xi\to\infty$, see Fig. \ref{fig1}. For small $\xi$, the asymptotic behavior of the probability is $P(\xi)\sim\xi$, indicating the increasing disturbance of the particle by the measurement apparatus. In the actual experiment \cite{S69}\cite{Le69}\cite{Z03}, the momentum precision $\Delta k$ is sometimes chosen twice the value of the first interference minimum (FIM), or equal to the full width at the half maximum (FWHM). According to (\ref{psip}), the momentum precision corresponding to the FIM is obtained by $\Delta k=2h/\Delta q$, which entails a probability $P(2)\approx 0.9$. Less significant is the probability of $P(0.89)\approx 0.72$ corresponding to the case of the FWHM with higher precision $\Delta k=0.89h/\Delta q$.
\\

In the following, we apply the concept of the 'measurement precision' in \cite{BGL95}\cite{D76}\cite{H82}, and consider the general conditional probability ${\cal P}_{k,q}(\Delta k\,|\,\Delta q;\psi)$ to measure the momentum $k$ of a particle with precision $\Delta k$, after having made a position selection at $q$ with the precision $\Delta q$. For every given measurement precisions $\Delta q$ and $\Delta k$ we will determine the {\it least upper bound} of ${\cal P}_{k,q}(\Delta k\,|\,\Delta q;\psi)$ by considering a variation problem in Hilbert space. \\

To start with, we consider a single particle in one spatial dimension described by a state vector, or wave function $\psi$ which is an element of the Hilbert space ${\cal H}=L^2(\mathbb{R})$, the space of square integrable functions on $\mathbb{R}$. We write $\hat\rho=|\psi\rangle\langle\psi|$ for the pure state in question. The scalar product in Hilbert space will be denoted by angular brackets, that is to write $\langle \phi|\psi\rangle$ for the scalar product of two state vectors $\phi,\psi\in{\cal H}$. Accordingly, the norm of $\phi$ is given by $||\psi||\equiv \sqrt{\langle \psi|\psi\rangle}$. Position and momentum of the system are represented as the Schr\"odinger pair of Operators $\hat x$, $\hat p$, where $(\hat x\, \psi)(x)=x\, \psi(x)$ and $(\hat p\, \psi)(x)=-i\hbar\, \psi'(x)$. \\

Let the vicinity $A_q\subset\mathbb{R}$ of a position value $q$ be defined by the half-open interval
$A_q=\big(q-\frac{\Delta q}{2},q+\frac{\Delta q}{2}\,\big]$, and let the vicinity $B_k\subset\mathbb{R}$ of a momentum value $k$ be defined by $B_k=\big(k-\frac{\Delta k}{2},k+\frac{\Delta k}{2}\,\big]$. Under a {\it projective position measurement} \cite{BGL95}\cite{D76}, performed on a state $\hat\rho$, the probability to measure the position $x\in A_q$ with precision $\Delta q$ has the form: $\mbox{tr}[\,\hat\rho\, E_{\hat x}(A_q)]=||E_{\hat x}(A_q)\,\psi||^2=\int_{A_q}|\psi(x)|^2 dx$, where $E_{\hat x}(A_q)$ is the value of the spectral measure or the positive operator-valued measure $E_{\hat x}$ on the vicinity $A_q\subset\mathbb{R}$ of $q$. Similar, the probability of $p\in B_k$ with the precision $\Delta k$ is given by $\mbox{tr}[\,\hat\rho\, E_{\hat p}(B_k)]$ where $E_{\hat p}(B_k)$ is the value of the spectral measure $E_{\hat p}$ on the vicinity $B_k\subset\mathbb{R}$ of $k$. In this case we have
$\mbox{tr}[\,\hat\rho\, E_{\hat p}(B_k)]=||E_{\hat p}(B_k)\,\psi||^2=\int_{B_k}|\tilde\psi(p)|^2 dp$
where $\tilde\psi$ is the Fourier transform of $\psi$. \\

Furthermore, the formalism for {\it conditional probabilities} under quantum measurements is very well developed \cite{BGL95}\cite{D76}\cite{H82}. In the initial measurement of the position, one may suppose either that the particle is absorbed during the measurement, or that it emerges in a state perturbed by the measurement. In the second case the uncertainty principle suggests that the more accurately the position is measured the greater is the perturbation of the momentum of the outgoing state, and there is no canonical instrument appropriate to this situation. A conventional way of treating this problem is to partition the position space into a countable number of disjoint sets, i.e. in the case considered above, $\{A_{q_i}\}$, $q_i=i\Delta q$, $i\in\mathbb{Z}$ and to take the outgoing state to be $\rho'= E_{\hat x}(A_{q_i})\,\rho\, E_{\hat x}(A_{q_i})$. By introducing another countable number of disjoint sets $\{B_{k_j}\}$, $k_j=j\Delta k$, $j\in\mathbb{Z}$, corresponding to the momentum measurement, the above mentioned conditional probability ${\cal P}_{k,q}(\Delta k\,|\,\Delta q;\psi)$ of a successful momentum measurement $p\in B_k$, {\it given} a previous position selection $x\in A_q$, is
\begin{eqnarray}\label{prob}
{\cal P}_{k,q}(\Delta k\,|\,\Delta q;\psi)=\frac{||E_{\hat p}(B_k)E_{\hat x}(A_q)\,\psi||^2}{||E_{\hat x}(A_q)\,\psi||^2}
\end{eqnarray}
For simplicity we suppressed the indices $i$ and $j$. Now, our main statement is the following:\\
\\
{\bf Theorem.} Let $\Delta q$ and $\Delta k$ be fixed. For every $q,k$ and $\psi\in{\cal H}$, the {\it least upper bound} of the measurement probability is given by the inequality
\begin{eqnarray} \label{ineq}
{\cal P}_{k,q}(\Delta k\,|\,\Delta q;\psi)\leq \xi\,\left[R^{(1)}_{00}(\pi\xi/2,1)\right]^2
\end{eqnarray}
with $\xi=\frac{\Delta k\Delta q}{h}$, and $R^{(1)}_{mn}(c,x)$ is the radial prolate spheroidal function of the first kind.\footnote{For the definition of $R^{(1)}_{mn}(c,x)$ see \cite{AS}. An extensive discussion of this special function can be found in \cite{M54}\cite{Str56}\cite{F57}.} \\
\\
{\bf Proof.} We reformulate (\ref{prob}) in order to be able to apply the subspace ${\cal H}_q=E_{\hat x}(A_q){\cal H}\subset {\cal H}$, equipped with the scalar product
\begin{eqnarray}\label{scalar}
\langle\phi|\psi\rangle_q=\int_{A_q}\phi^*(x)\,\psi(x)\,dx
\end{eqnarray}
and norm $||\,\psi||_q=\sqrt{\langle\psi |\,\psi \rangle_q}$. Initially, we consider the linear mapping $\hat{G}_{kq}:{\cal H}_q \to {\cal H}_q$, defined by
\begin{eqnarray}\label{G}
(\hat{G}_{kq}\psi)(x)=\int_{A_q}\; g_k(x-x')\;\psi(x')\;dx'
\end{eqnarray}
with the convolution kernel
\begin{eqnarray}\label{g}
g_k(x)= e^{\frac{i}{\hbar}k\,x}\;\frac{\sin(\frac{\Delta k}{2\hbar}\,x)}{\pi x}
\end{eqnarray}
This kernel is continuous, bounded and $g_k(x)=g^*_k(-x)$, i.e. the operator $\hat{G}_{kq}$ is self-adjoint. Then, we obtain the following representation of (\ref{prob})
\begin{eqnarray}\label{PG}
{\cal P}_{k,q}(\Delta k\,|\,\Delta q;\psi)
=\frac{\langle\psi|\,\hat{G}_{kq}\,\psi\rangle_q}{\langle\psi|\psi\rangle_q}
\end{eqnarray}
On the other hand, the operator norm of $\hat{G}_{kq}$ in ${\cal H}_q$ is formally given by
\begin{eqnarray} \label{Norm}
||\hat{G}_{kq}||_q=\sup\limits_{\psi\in {\cal H}\setminus\{0\}}\frac{|\langle\psi|\,\hat{G}_{kq}\,\psi\rangle_q|}{\langle\psi|\psi\rangle_q}
\end{eqnarray}
and simply obtains the least upper bound of the measurement probability (\ref{prob}). A substantial step for the computation of $||\hat{G}_{kq}||_q$ is given by the following: \\
\\
{\bf Lemma.} For every $q,k, \Delta q$ and $\Delta k$, we receive the identity
\begin{eqnarray} \label{lem}
||\hat{G}_{kq}||_q=||\hat{G}_{00}||_0
\end{eqnarray}
{\bf Proof.} We consider the translation $\hat{T}_{q}$ defined by $(\hat{T}_{q}\psi)(x)=\psi(x-q)$ and the unitary transformation $\hat{U}_{k}$ with $(\hat{U}_{k}\psi)(x)=e^{\frac{i}{\hbar} k\, x}\psi(x)$. Then, by using the identities
\begin{eqnarray} \label{ident}
\langle\psi|\,\hat{G}_{kq}\,\psi\rangle_q&=&\langle\varphi_{kq}|\,\hat{G}_{00}\,\varphi_{kq}\rangle_0\\
\langle\psi|\,\psi\rangle_q &=&\langle\varphi_{kq}|\,\varphi_{kq}\rangle_0
\end{eqnarray}
with $\varphi_{kq}=(\hat{U}_{k}\hat{T}_{q})^{-1}\psi$, there is the following reformulation of (\ref{Norm})
\begin{eqnarray} \label{Normq}
||\hat{G}_{kq}||_q = \sup\limits_{\varphi\in (\hat{U}_{k}\hat{T}_{q})^{-1}{\cal H}\setminus\{0\}}
\frac{|\langle\varphi|\,\hat{G}_{00}\,\varphi\rangle_0|}{\langle\varphi|\varphi\rangle_0}
\end{eqnarray}
By using ${\cal H}=\hat{U}_{k}\hat{T}_{q}{\cal H}$ the lemma is proven.\\
\\
Now, as $\hat{G}_{00}$ is a compact and self-adjoint linear operator, there is a real eigenvalue with modulus equal to $||\,\hat{G}_{00}||_0$. It is easy to show that $\hat{G}_{00}$ is positive definite on ${\cal H}_0$ and $||\,\hat{G}_{00}||_0$ is equal to the maximal eigenvalue of $\hat{G}_{00}$. According to (\ref{G}) and (\ref{g}), the eigenvalues of $\hat{G}_{00}$ must satisfy the following homogeneous Fredholm integral equation of the second kind
\begin{eqnarray} \label{eig}
\lambda_n\,\psi_n(x)=\frac{1}{\pi}\,\int_{-1}^{1}\frac{\sin(\frac{\pi}{2}\xi(x-y))}{x-y}\;\psi_n(y)\,dy 
\end{eqnarray}
with $|x|\leq 1$, and the single parameter, $\xi$, appears instead of $\Delta q$ and $\Delta k$ separately. From standard theory we know that (\ref{eig}) has solutions in $L^2([-1,1])$ only for a discrete set of eigenvalues, $\lambda_0\geq\lambda_1\geq,...$ and that as $n\to\infty$, $\lim\lambda_n\to 0$. It should be noted that both the $\psi_n(x)$ and $\lambda_n$ depend on the parameter $\xi$. A detailed mathematical analysis of equation (\ref{eig}), and some asymptotic expansions for prolate spheroidal wave functions are given in \cite{S65}. Corresponding to each eigenvalue $\lambda_n(\xi)$ there is a unique solution $\psi_n(x)=S_{0n}(\pi\xi/2,x)$ called {\it angular prolate spheroidal wave function}.\footnote{A number of books \cite{M54}\cite{Str56}\cite{F57} treat the prolate wave functions in detail.} They are continuous functions of $\xi$ for $\xi\geq 0$, and are orthogonal in $(-1,1)$. Moreover, they are complete in $L^2([-1,1])$. The corresponding eigenvalues are related to a second set of functions called {\it radial prolate spheroidal functions}, which differ from the angular functions only by a real scale factor.
Applying the notation of Flammer \cite{F57} the eigenvalues are
\begin{eqnarray} \label{values}
\lambda_n(\xi)=\xi\,\left[R^{(1)}_{0n}(\pi\xi/2,1)\right]^{\,2}
\end{eqnarray}
with $n=0,1,2,...$ These eigenvalues are non-degenerate for $\xi>0$ and one can prove that $\lambda_0>\lambda_1>...> 0$. Thus, the largest eigenvalue is $\lambda_0(\xi)$ and we obtain
\begin{eqnarray} \label{nq}
||\hat{G}_{00}||_0=\lambda_0(\xi)
\end{eqnarray}
corresponding to the statement of the theorem.$\mbox{ }$ $\hfill$ $\Box$\\

Various algorithms for the numerical computation of the prolate spheroidal functions are discussed in \cite{WS05}\cite{LW98}. Most of the standard methods involve an expansion of Legendre polynomials for small values and expansion in Bessel functions in the neighborhood of infinity. In Fig. \ref{fig1}, we see the monotonically increasing behavior of $\lambda_0(\xi)$. For small values of $\xi$, the behavior of $\lambda_0(\xi)$ is given by
\begin{eqnarray} \label{small}
\lambda_0(\xi)=\xi\,\left[1-\left(\frac{\pi\xi}{6}\right)^2+{\cal O}(\xi^4)\right]
\end{eqnarray}
with $\lambda_0(\xi)\sim \xi$ for $\xi\to 0$. Actually, the leading term of this expansion is equal to the trace of $\hat{G}_{kq}$, which is, according to Mercer's theorem, given by
\begin{eqnarray} \label{tr}
\mbox{Tr}(\,\hat{G}_{kq})=\xi
\end{eqnarray}
and $\lambda_0(\xi)$ can never exceed the trace. An alternative upper bound of $\lambda_0(\xi)$ is obtained by the Hilbert-Schmidt-norm of $\hat{G}_{kq}$. The computation is straightforward by applying the ordinary integral representation
\begin{eqnarray} \label{GHS}
||\hat{G}_{kq}||_{HS}=\big[\int_{A_q}\int_{A_q}\!\! |\,g_k(x-x')|^2\;dx\;dx'\;\big]^\frac{1}{2}
\end{eqnarray}
and according to (\ref{g}) we immediately obtain the expression \footnote{The sine-integral is Si$(x)=\int_0^x\frac{\sin(t)}{t}\,dt$ respectively Cin$(x)=\int_0^x\frac{1-\cos(t)}{t}\,dt$, see \cite{AS}.}
\begin{eqnarray}\label{GHS1}
||\hat{G}_{kq}||_{HS}&=&\frac{1}{\pi}\,\Big[\,2\pi\xi\;\mbox{Si}(2\pi\xi)-\mbox{Cin}(2\pi\xi)\nonumber\\
& &+\cos(2\pi\xi)-1\,\Big]^\frac{1}{2}
\end{eqnarray}
This bound is slightly tighter than the trace, and it is non-trivial for $\xi\leq 1.37$. Instead, for large values of $\xi$ an asymptotic expansion of $\lambda_0(\xi)$ is given by the following expression \cite{F64}
\begin{eqnarray} \label{large}
\lambda_0(\xi)=1-\pi\sqrt{8\xi}\;e^{-\pi\xi}\left[1-\frac{3\pi}{64}\,\xi+{\cal O}(\xi^{-2})\right]
\end{eqnarray}
whereas the convergence behavior is mainly determined by the exponential damping factor \footnote{The area between $\lambda_0(\xi)$ and $1$ is finite and we numerically obtain the value $\int_0^\infty(1-\lambda_0(\xi))\,d\xi=0.65077(5)$.}. \\

On the other hand, empirically we found that the function $\text{erf}(\frac{\sqrt{\pi}}{2}\,\xi)$ is proceeding slightly above $\lambda_0(\xi)$, as we can see in Fig. \ref{fig1}.
Moreover, it preserves the property to vanish for $\xi=0$ with slope $1$, and it is monotonically increasing with an upper bound of $1$. Numerically we found, that the maximum of the deviation from $\lambda_0(\xi)$ is less than $1\%$ and is localized in the neighborhood of $\xi\approx 1.48$. We have not been able to falsify the inequality $\lambda_0(\xi)\leq\text{erf}(\frac{\sqrt{\pi}}{2}\,\xi)$ and thus conjecture it to be a proper upper bound for all $\xi\geq 0$.

The vertical line $\xi=1$ in Fig. \ref{fig1} is the ordinary dividing line ('unit step') of Heisenberg corresponding to the relation (\ref{H2}). Instead, according to the least upper bound $\lambda_0(\xi)$, we additionally consider probabilistic aspects of the measurement process. Consequently, no measurement event with conditional probability above $\lambda_0(\xi)$ does exist. According to the monotonic behavior of $\lambda_0(\xi)$, such an exclusion occurs for both $\xi<1$ and $\xi\geq 1$. For instance, measurement events with precisions $\Delta k\Delta q=h$ and probabilities greater than $\lambda_0(1)= 0.78$ are impossible \footnote{This value might be a hint for the necessity of the notation "$\sim$" in Heisenberg's original inequality (\ref{H2}).}. Furthermore, for precisions with $\Delta k\Delta q=\hbar=h/2\pi$, as applied in the textbook of Landau and Lifschitz (\cite{LL}, p. 45), the least upper bound of the measurement probability is merely $\lambda_0(\frac{1}{2\pi})= 0.16$. In fact, for the constitution of a proper measurement apparatus, higher values of $\lambda_0(\xi)$ should be preferred, e.g. a bound $\lambda_0(\xi)\geq 0.98$ is corresponding to the necessary condition $\Delta k\Delta q\geq 2h$.  \\

The case of minimum uncertainty in (\ref{K}) is achieved for Gaussian state functions saturating the lower limit of the ordinary uncertainty principle, i.e. $\sigma_x\sigma_p=\hbar/2$. According to our theorem, the bound $\lambda_0(\xi)$ can not be attained by the measurement probability (\ref{prob}) in this case. Instead, it is reached for the {\it prolate angular spheroidal} eigenfunction, $\psi_0(x)=S^{(1)}_{00}(\frac{\pi\,\xi}{2},x)$, corresponding to the maximum eigenvalue $\lambda_0(\xi)$ (see theorem). \\

Actually, the least upper bound is just as valid for measuring processes which are carried out in reversed order. We obtain the corresponding conditional probability by the change of the projectors $E_{\hat{x}}(A_q)$ and $E_{\hat{p}}(B_k)$ in (\ref{prob}). Then, the derivation is done in the momentum representation and is identical with the original derivation in the position representation, except for the sign of the imaginary unit. Due to the independence of the norm of $q$ and $k$ (see lemma), the bounds are same as before. \\

Furthermore, a generalization of our results to consecutive position measurements with finite time-delay is possible. In this case we consider two successive position measurements at $q$ and $q'$ with time-delay $t>0$, and the corresponding precisions are $\Delta q$ and $\Delta q'$. In analogy to our lemma, the norm of the appropriate operator is independent of $q$ and $q'$. Therefore, we obtain the same bounds as before except that we have to replace the parameter $\xi$ by $\tilde\xi=\frac{m}{t}\frac{\Delta q\Delta q'}{h}$ in (\ref{values}) and (\ref{nq}), where $m$ is the mass of the particle. The latter might be interesting as spin-measurements in the Stern-Gerlach experiment are principally produced by two consecutively position measurement. In this case, $\Delta q$ corresponds to the gap of the pols of the magnet where the particle emerges from, and $\Delta q'$ is given by the domain of the screen where the spin of the particle is red as 'up' or 'down'. But if the time interval $t$ of the two measuring events is so big that the inequality $\tilde\xi\ll 1$ is valid, this is a clear indication that there is an essential disturbance of the measurement result caused by the measurement device. On the other hand, too small values of $t$ might lead to the problem, that no sufficient separation between the two spin directions is produced. Therefore, it seems interesting to reexamine these experiments in more detail.  \\

In summary, we considered Heisenberg's concern to establish a quantitative expression for the minimum amount of unavoidable momentum disturbance caused by any position measurement. We proposed to apply the conditional probability of consecutive position and momentum measurements. As our main result, we derived a tight upper bound of this probability. This bound is independent of the state vector, and is just as valid for measuring processes which are carried out in reversed order.


\newpage{}
\end{document}